# Title: Observation of room-temperature ferroelectricity in elemental Te nanowires

**Authors:** Jinlei Zhang[1,#], Jiayong Zhang[1,#], Yaping Qi[2,#], Shuainan Gong[1,#], Run Zhao[1], Hongbin Yang[3], Zhenping Wu[4], Dapeng Cui[5], Lin Wang[6], Chunlan Ma[1], Ju Gao[1,7], Yong P. Chen[8,9,2] * and Yucheng Jiang[1*]

**Affiliations:**

[1]Jiangsu Key Laboratory of Micro and Nano Heat Fluid Flow Technology and Energy Application, School of Physical Science and Technology, Suzhou University of Science and Technology, Suzhou, 215009, China

[2]Department of Engineering Science, Faculty of Innovation Engineering, Macau University of Science and Technology, Av. Wai Long, Macau SAR, 999078, China

[3]Institute of Materials Science & Devices, Suzhou University of Science and Technology, Suzhou 215009

[4]State Key Laboratory of Information Photonics and Optical Communications & School of Science, Beijing University of Posts and Telecommunications, Beijing 100876, China

[5]Department of Physics, Faculty of Science, National University of Singapore, Singapore 117551, Singapore

[6]School of Materials Science and Engineering, Shanghai University, Shanghai 200444, China

[7]School for Optoelectronic Engineering, Zaozhuang University, Shandong 277160, China

[8]Department of Physics and Astronomy and Elmore Family School of Electrical and Computer Engineering and Birck Nanotechnology Center and Purdue Quantum Science and Engineering Institute, Purdue University, West Lafayette, Indiana 47907, United States

[9]Institute of Physics and Astronomy and Villum Center for Hybrid Quantum Materials and Devices, Aarhus University, Aarhus-C, 8000 Denmark

[#]These authors contributed equally: Jinlei Zhang, Jiayong Zhang, Yaping Qi, Shuainan Gong.

*Corresponding authors. Email: yongchen@purdue.edu (Y.P.C.) and jyc@usts.edu.cn (Y.C.J.).

**Abstract:** Ferroelectrics are essential in low-dimensional memory devices for multi-bit storage and high-density integration. A polar structure is a necessary premise for ferroelectricity, mainly existing in compounds. However, it is usually rare in elemental materials, causing a lack of spontaneous electric polarization. Here, we report an unexpected room-temperature ferroelectricity in few-chain Te nanowires. Out-of-plane ferroelectric loops and domain reversal are observed by piezoresponse force microscopy. Through density functional theory, we attribute the ferroelectricity to the ion-displacement created by the interlayer interaction between lone pair electrons. Ferroelectric polarization can induce a strong field effect on the transport along the Te chain, supporting a self-gated field-effect transistor. It enables a nonvolatile memory with high in-plane mobility, zero supply voltage, multilevel resistive states, and a high on/off ratio. Our work provides new opportunities for elemental ferroelectrics with polar structures and paves a way towards applications such as low-power dissipation electronics and computing-in-memory devices.

**One-Sentence Summary:** We find out-of-plane ferroelectricity in elemental Te nanowire, allowing a self-gated ferroelectric field-effect transistor.

Low-dimensional ferroelectrics have drawn extensive interest in nanoelectronics due to their charge polarizations tunable by external electric fields, for applications as transistors, memories, and sensors (*1*). The direction of the spontaneous polarization is important to determine possible applications. The out-of-plane ferroelectricity is preferred to realize memory miniaturization with high integration density and low power consumption (*1-4*), compared with the in-plane one. Usually, ferroelectrics are insulators with low carrier mobility and large band gaps (*1-6*). For functional nanodevices, an effective strategy is to construct heterostructures by combining ferroelectric gate insulators and channel semiconductors (*7-13*) to make ferroelectric field-effect transistors (Fe-FETs). However, leakage current and charge trapping usually exist at the interface, limiting application prospects. A material that is both a high-mobility in-plane semiconductor and out-of-plane ferroelectric could make an ideal memory nanodevice and overcome the above disadvantages (*14-17*). Such a material requires a highly anisotropic lattice structure with different in-plane and out-of-plane properties. Recently, a two-dimensional (2D) ferroelectric semiconductor α-$In_2Se_3$ has been used as the channel material to compose a Fe-FET (*14*). The spontaneous out-of-plane polarization can tune the in-plane electronic transport via a gate voltage to give rise to a nonvolatile resistive switching, achieving a self-gated Fe-FET (SF-FET). Until now, many efforts have been made to explore ferroelectric semiconductors, from two-dimensional materials (such as α-$In_2Se_3$) (*14-15, 18*) to emerging moiré superlattices (such as twisted $MoS_2$ bilayers) (*16, 19-20*).

Ferroelectricity is mainly found in compounds with non-centrosymmetric structures (*15, 19-25*), which are unexpected in elemental crystals. Generally, elemental materials possess centrosymmetric structures without local electronegativity differences, thereby lacking spontaneous electric polarizations (*25-27*). To date, it has still been a challenge to realize elemental ferroelectrics. Recently, tellurium (Te) has drawn significant interest as a 2D elemental semiconductor (*28*), due to its outstanding electrical conductivity (*29-30*), one-dimensional chain structure (*31-32*), and strong spin-orbit coupling (SOC) (*33-34*). Theories have predicted that layered Te possesses a broken symmetry and can be a ferroelectric semiconductor (*27*). The interlayer interaction between adjacent Te chains drives the atomic displacement accompanied by charge transfer, separating the charge centers and leading to spontaneous electric polarization. However, until now, there has been no experimental evidence to support such a prediction. Most recent studies have focused on the Te nanosheets, whose highly stable structure may limit the reversal of dipoles. Another form of Te is the nanowire with a relatively low barrier height of atomic displacement (*35-36*), providing a possible route to achieve ferroelectricity.

In this work, we report room-temperature out-of-plane ferroelectricity in few-chain Te nanowires and the edges of Te nanosheets. The ferroelectric hysteresis and domain switching are observed by piezoresponse force microscopy (PFM). Using spherical aberration corrected scanning transmission electron microscopy (Cs-corrected STEM), we attribute the out-of-plane ferroelectricity to the atomic displacement perpendicular to the atomic chains in Te nanowires, verified by density functional theory (DFT) calculations. Also, the Te nanowire can be used as a ferroelectric semiconductor channel to construct a SF-FET with a *h*-BN gate dielectric. It shows a nonvolatile resistive memory effect at zero gate supply voltage, with a large threshold voltage window, high on/off ratio, long-term retention, and multilevel resistive states.

Few-chain Te nanowires were synthesized with a width of 30~300 nm and a height of 5~30 nm using a substrate-free solution method (Materials and Methods and fig. S1). The surface morphology and cross-section of a representative Te nanowire probed through atomic force microscopy (AFM) and scanning electron microscopy (SEM) are shown in Fig. 1A. It shows a triangle-like cross section with clean surfaces and neat edges, indicating the high crystal quality. The STEM image shows a unique chiral chain lattice in Fig. 1B. Each Te atom is bonded to two nearest neighbors, forming helical chains along the *c*-axis direction (which is the long axis of the nanowire). Those chains are stacked in a hexagonal array through van der Waals bonds between adjacent atomic chains. Note that the Te nanowire shows an inclined plane along the *b*-axis at the edge. It can be thought to be formed by multiple Te atomic layers stacked along the $\vec{e}$ direction (perpendicular to *b-c* plane). Layer-center atoms (LCAs) in each single Te chain show relative off-center displacements, thereby giving rise to a non-centrosymmetric lattice structure of the inclined plane at the edge (see the right in Fig. 1B). As is well known, the breaking of symmetry is an important premise of ferroelectricity.

Considering the experimental case, we used DFT to simulate the ferroelectric characteristics of few-layer Te, focusing on modeling the off-center displacement of LCAs at the inclined edge (which as discussed below would give an out-of-plane polarization component measured in our experiment). To model the layers near and parallel to the inclined edge and stacked along the $\vec{e}$ direction of the Te nanowire, we performed detailed calculations for a 4-layer (4L) Te sheet in three different phases. The atomic structures of the *α*-, *β*-, and *α′*-phase 4L Te are plotted in the left, middle, and right insets in Fig. 1C, respectively. The *β*-phase Te has the space group $P2_1/m$ and is centrosymmetric, while the space group of the *α* (*α′*)-phase Te is reduced to $P2_1$. The *α*-phase and *α′*-phase Te are energetically degenerate with the opposite displacement (about ± 0.41 Å) of their LCAs. The climbing image nudged elastic band (NEB) calculations in Fig. 1C show that the two stable structures of non-centrosymmetric *α*- and *α′*-phase are connected through a saddle point with the centrosymmetric *β*-phase, implying the possible existence of ferroelectricity. The calculated ferroelectric switching pathway of 4L Te from *α*- to *α′*-phase in Fig. 1C shows that the energy barrier between the non-centrosymmetric and centrosymmetric phases is about 67 meV per layer. We also calculated energy barriers for different numbers of Te layers (fig. S3). The differential charge densities are calculated to investigate the physical mechanism of the existence of polarization in the *α*- and *α′*-phase Te (see Fig. 1D), with the red and blue colors denoting the electron density charge accumulation and depletion regions, respectively. For the *β*-phase 4L Te, the charge distribution is centrosymmetric around the LCAs, resulting in the positive and negative charge centers canceling each other to give a zero net polarization. In contrast, the differential charge densities of the α- and α′-phases show charge redistributions between the LCAs and their neighboring atoms. For the α- (or α′-) phase layered Te, the negative charge center moves to the right (or left) side of the positive charge center due to more charge accumulating at the right (or left), resulting in the emergence of nonzero polarization pointing from right (or left) to left (or right). Fig. 1E shows the calculated spontaneous polarization per layer as a function of the number of layers for α-phase Te. We do not calculate a monolayer as it has only the β-phase as the stable state and it alone cannot make a meaningful nanowire with inclined edges *(27)*. The 2D polarization density is about (2.1~2.4) × 10$^{-10}$ C/m per layer and slightly increases with increasing number of layers below 8 layers. Apparently, the polarization starts to decrease with further increasing number of layers beyond 8 layers. All of our calculations strongly suggest the existence of ferroelectricity within the layer (perpendicular to the Te chains), thus along the *b*-axis of the elemental Te nanowire giving rise to an out-of-plane (perpendicular to the *a-c* plane and substrate) polarization component. Note our few-layer model can also apply to the Te layers parallel to the $\vec{a}$ direction (e.g., near the bottom surface of the Te nanowire), implying an in-plane polarization (along the *a*-axis) can also exist. This however is not the focus of our experiment, which mainly probes the out-of-plane polarization.

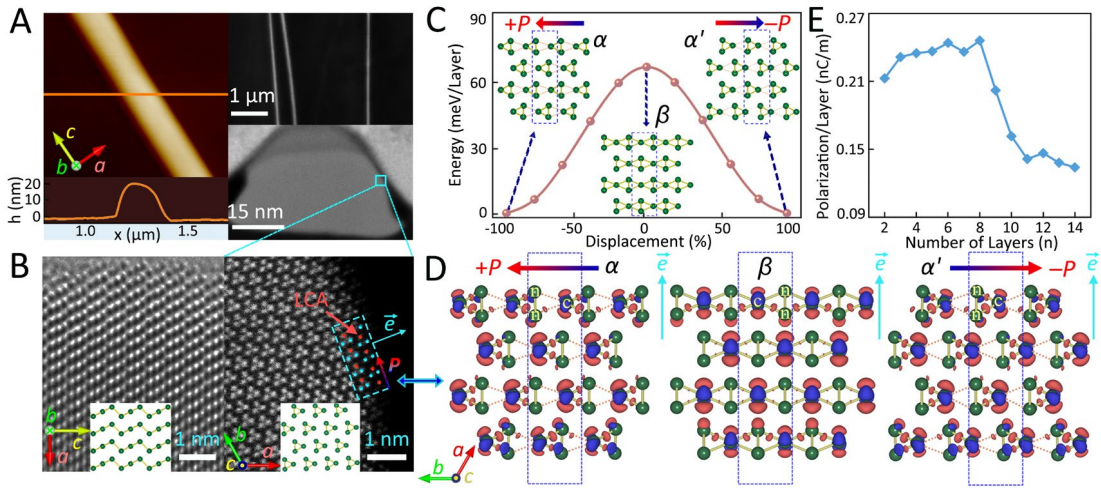

**Fig. 1. Polar structure and ferroelectric polarization.** (**A**) Clockwise from upper left: AFM, SEM, TEM topographical images and the AFM height profile of a Te nanowire on a SiO$_2$/Si wafer. (**B**) STEM images of the Te nanowire viewed along *b*- and *c*-axes (lower left insets here as well as in (A) show axes projected into the image planes; left and right insets show atomic lattice models). $\vec{e}$ is the epitaxial direction of inclined stepwise planes. (**C**) Ferroelectric switching pathway of 4L Te from α- to α′-phase calculated using the NEB method. Insets show atomic structures of α and α′ ferroelectric phases and centrosymmetric nonpolar β-phase. (**D**) Differential charge density (Δρ) of 4L Te for α-, β- and α′-phases. Here Δρ = ρ(tot) – ρ(c-Te) – ρ(n-Te), where ρ(tot) is the total charge density, ρ(c-Te) is the charge density of the LCAs (c) and ρ(n-Te) is the charge density of the neighboring (n) Te atoms. Red and blue colors represent the charge accumulation and depletion, respectively. The isosurface values in α- (α′-) and β-phases are set to 0.007 and 0.006 e/bohr$^3$, respectively. The arrows denote the polarization direction. Dotted rectangular boxes in **C** and **D**

indicate the smallest repeatable unit in the periodic lattice. Dotted rectangular boxes in **C** and **D** indicate the smallest repeatable unit in the periodic lattice. (**E**) Calculated polarization per layer for the α-phase Te with different number of layers.

Out-of-plane PFM was performed to investigate the spontaneous electric polarization of the Te nanowire. Te nanowires were transferred onto an Au-coated silicon substrate for the PFM measurement. Fig. 2A, measured on a representative Te nanowire (width ~300 nm, thickness ~18.9 nm), shows that the piezoresponse peaks at a resonance frequency of ~ 365 kHz, whose intensity is linear with the dc bias voltage (see the inset in Fig. 2A) applied to the tip. It verifies the existence of electric dipole moment and piezoelectricity with an effective piezoelectric coefficient $d_{33}$ of ~3.1 pm/V in the Te nanowire. Both phase hysteresis and butterfly amplitude loops, measured by sweeping the dc bias voltage applied to the tip, are observed in Figs. 2 B and C, respectively. It is evident that the electric polarization can be switched by dc bias poling, indicating a ferroelectric behavior. For comparison, Te nanowires were also transferred to an N-doped conductive silicon substrate to exclude the possible effect of substrates, showing a similar ferroelectric hysteresis. In Fig. 2D, the PFM is used to identify the distribution of polarization within the Te nanowire, reflecting its local ferroelectric domain structure. The phase and amplitude mappings correspond well to each other, overlaid on the three-dimensional (3D) topography (see Fig. 2D and fig. S4). The ferroelectric domain pattern shows a strong contrast between upward (orange) and downward (olive) polarizations with domain walls along the chain direction in the Te nanowire. Such a domain structure can be attributed to the opposite off-center ion displacements between the two inclined edges (see fig. S5). The narrow olive belt at the right edge may result from a slight bend at the end of the edge (see Fig. 1A, lower right). For a ferroelectric, off-center ion displacements are tunable by mechanical or electric driving, thereby reversing the ferroelectric domains. Here, the reversal of domains is evaluated after electrical poling using a dc bias applied to the proximal tip. We investigate the reversal of the ferroelectric domain in the Te nanowire by "poling" with opposite dc bias (±2.5 V) applied in Figs. 2 E and F. The orientation of ferroelectric polarization is switchable with the external electric field. Our theoretical and experimental results demonstrate that the ion displacements at the inclined edge give rise to an out-of-plane component of electric polarization that can be locally detected and controlled by an external field.

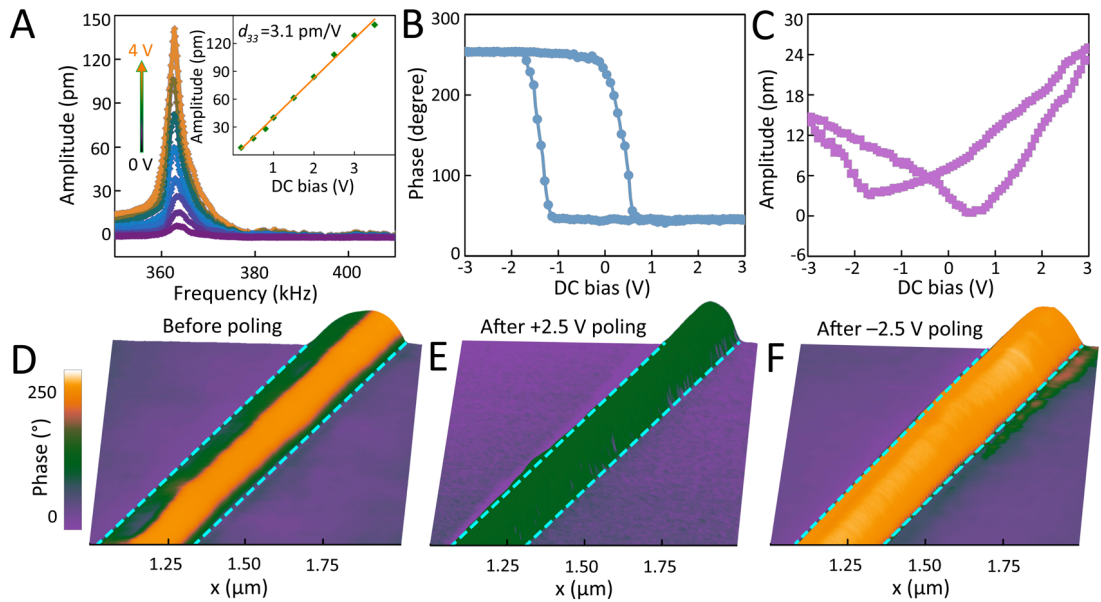

**Fig. 2. Ferroelectric hysteresis and domain switching.** (**A**) Piezoresponse vs. frequency under different dc bias voltages, measured at a representative location on the inclination edge and showing a resonance peak around 365 kHz. Inset shows the linear dependence of piezoresponse amplitude on dc bias with an effective piezoelectric coefficient $d_{33}$ of ~3.1 pm/V. (**B** and **C**) Phase-voltage hysteresis loop (B) and amplitude-voltage butterfly loop (C) measured by vertical PFM. (**D**, **E** and **F**) Phase mapping measured by vertical PFM before electric poling (D), after poling at +2.5 V (E), and after poling at –2.5 V (F), overlaid on the 3D topography image of the nanowire. The area of Te nanowire is marked by the cyan dashed line.

The origin of the ferroelectricity in Te nanowires implies the importance of the inclined edge. Besides the Te nanowire, one can synthesize Te nanosheet by a similar substrate-free solution process. Therefore, it is intriguing to investigate the usually neglected edge ferroelectric responses of Te nanosheets. Fig. 3A shows the SEM and AFM images of a Te nanosheet with a width of ~1.2 μm and a thickness of ~15 nm. An inclined edge of the Te nanosheet is shown in Fig. 3B, providing a possible platform to investigate edge ferroelectricity. It is intriguing that the effective piezoelectric coefficients $d_{33}$ at the edge was found to be about 80% higher than that inside (Fig. 3B and fig. S6). The difference in the piezoelectric responses between the edge and inside is also reflected by the resonance frequencies (see fig. S6), which are about 371 kHz (edge) and 391 kHz (inside) respectively. We note that the existence of piezoelectric resonance itself does not necessarily imply ferroelectricity. The edge electric polarization of the Te nanosheet is also investigated using the out-of-plane PFM in Fig. 3C. The phase hysteresis loops were measured at various positions. The coercive field decreases significantly with the tip moving from the edge to the inside, indicating that the ferroelectricity emerges only at the edge of the Te nanosheet. In comparison, for our Te nanowires with a width of 20-300 nm, the top surface is full of inclined steps, causing out-of-plane ferroelectricity in all regions. To control the ferroelectric domain near the edge of the Te nanosheet, we used a biased tip above the local domain to apply an electric field while scanning the Te nanosheet. The PFM phase images are overlaid on the 3D topography of the nanosheet after dc bias poling in Figs. 3, D to F. We observe the post-polling shift of the domain wall, demonstrating the switching of ferroelectric polarization induced by the external electric field. The dark region, corresponding to the ferroelectric domain, tends to shrink after a positive poling of +2.5 V (see Fig. 3E), but expand after a negative poling (see Fig. 3F).

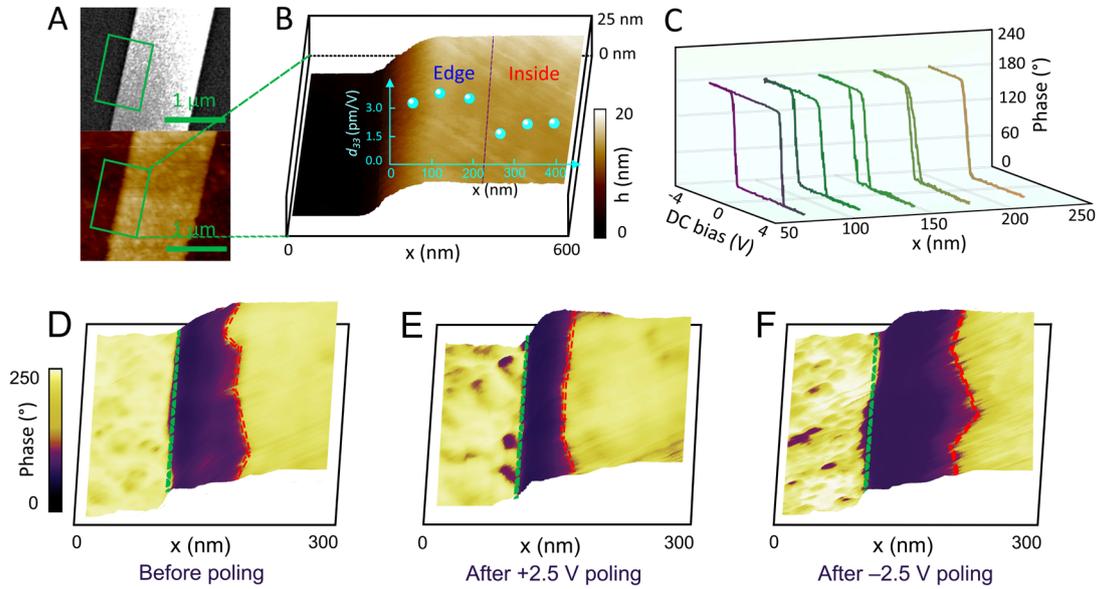

**Fig. 3. Edge ferroelectricity in Te nanosheet.** (**A**) Upper and Lower: SEM, and AFM topographical images of a Te nanosheet (height of ~ 15 nm) on an Au-coated silicon wafer. (**B**) Values of $d_{33}$ at the different marked positions (overlaid on an AFM topo-graph) of the Te nanosheet, showing that $d_{33}$ at the edge is about 80 % higher than that inside of the nanosheet. (**C**) PFM phase hysteresis loops from the edge to the inside of 2D Te nanosheet, showing ferroelectric domain with edge width less than 200 nm. (**D**, **E** and **F**) Vertical PFM phase mapping before electric poling (D), after poling at +2.5 V (E), after poling at –2.5 V (F), overlaid on 3D topography image of the Te nanosheet.

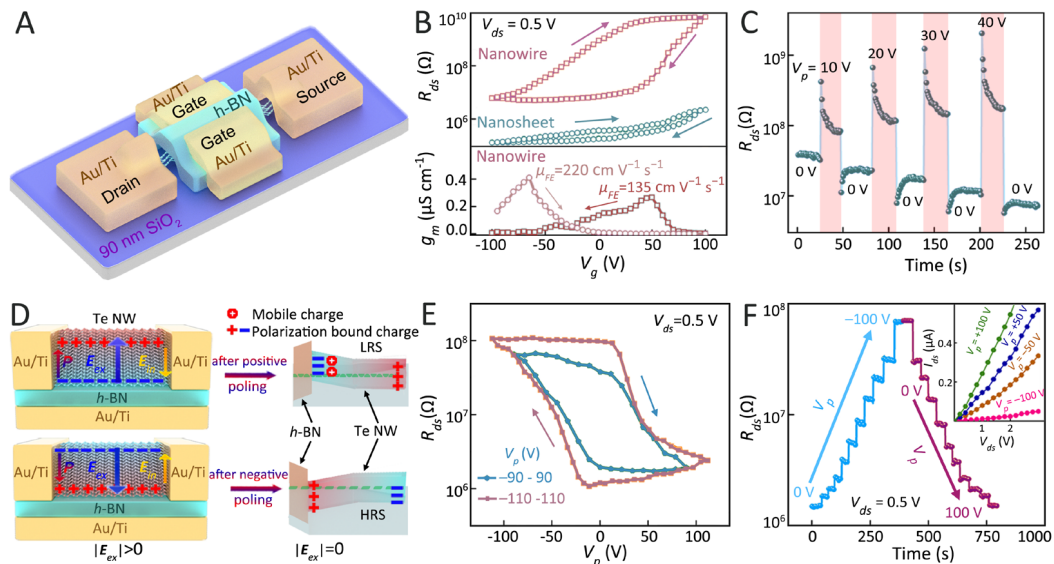

**Fig. 4. Switching characteristics exhibiting nonvolatile memory of a Te nanowire self-gated**

**ferroelectric field effect transistor (SF-FET). (A)** Schematic showing the device structure. **(B)** $R_{ds}$-$V_g$ characteristics for the Te nanowire and nanosheet SF-FETs (upper) and $g_m$–$V_g$ hysteresis loop of Te nanowire SF-FET (lower). **(C)** Nonvolatile switching of $R_{ds}$ measured after applying multiple $V_p$ square pulses with the pulse width of 25 s, exhibiting multilevel resistive states. **(D)** Schematic showing polarization bound charge distribution with an external electric field applied and band structures for LRS and HRS at zero field. **(E)** $R_{ds}$-$V_p$ hysteresis loops at $V_{ds}$= 0.5 V. Measurement of each data point is performed at $V_g$ = 0 V after waiting for 1 minute following applying each $V_p$ pulse of 100-ms width. **(F)** Nonvolatile resistive states switched by different $V_p$. Resistive states are measured after applying $V_p$ pulses (from –100 V to 100 V then back to –100 V in steps of 10 V) of 100-ms width. The inset is the linear $I_{ds}$-$V_{ds}$ characteristics after applying different $V_p$ pulses labeled.

Given that Te is a p-type semiconductor (see figs. S7 and S8) (*30, 32, 35*), a ferroelectric Te nanowire can be used as the channel material of a SF-FET via its out-of-plane spontaneous electric polarization. Fig. 4A shows a SF-FET device composed of a Te nanowire as the channel material and a *h*-BN layer as the top gate dielectric. It possesses a typical (p-type) field-effect behavior exhibited by the transfer curves in Fig. 4B and fig. S8, where a gate voltage ($V_g$) sweeps to vary the carrier density. As a reference for the transport characteristic of a transistor, effective extrinsic field-effect mobility ($\mu_{FE}$) values of ~135 and ~220 cm$^2$ V$^{-1}$ s$^{-1}$ for the backward and forward gate voltage sweeps can be extracted from the maximum transconductance ($g_m$) (Fig. 4B, lower), comparable to the high-mobility FETs based on few-layer black phosphorene (*37*) and few-layer MoS$_2$ (*38*). It is most intriguing to observe the resistance-$V_g$ ($R_{ds}$-$V_g$) hysteresis originating from the switching of out-of-plane electric polarization. This provides strong evidence to support the ferroelectricity of the Te nanowire, and suggests a potential application for resistive memories. For comparison, we have also investigated the field effect of Te nanosheet that shows a much smaller hysteresis behavior. The out-of-plane ferroelectricity exists only at the edge, so the spontaneous polarization is not large enough to gate the whole Te nanosheet. In addition, we note that both Te nanowire and nanosheet exhibit clockwise loops in $R_{ds}$-$V_g$, different from a regular Fe-FET (*10, 14*). For the self-gated device, the external electric field ($E_{ex}$) has an opposite direction to the internal remnant electric field ($E_{in}$) induced by the spontaneous polarization, thereby causing competition and balance between them. To clarify this hysteresis behavior, we studied the resistive response to multiple on/off operations of a pulsed gate voltage ($V_p$) as shown in Fig. 4C. It is observed that $R_{ds}$ increases sharply when $V_p$ is applied, but decrease to a level even lower than the initial resistance once $V_p$ is removed (so $V_g$ = 0 V again). More and even lower resistive states can be obtained at zero $V_g$ after more $V_p$ pulses with increasing amplitudes. Additional data after various pulsed voltage ($V_p$) can be found in figs. S9 and S10. To interpret this effect, a schematic diagram is shown to describe the gating process and the band structures for high resistive state (HRS) and low resistive state (LRS) in Fig. 4D. In the case of continuous application of positive $V_g$ or $E_{ex}$, the $R_{ds}$ increase is mainly determined by $E_{ex}$ because of $E_{ex} \gg E_{in}$. In the absence of $E_{ex}$, $E_{in}$ has an opposite direction to $E_{ex}$, thus has a similar effect to a negative $V_g$ or $E_{ex}$ that attracts more mobile holes to the (p-type) channel and gives rise to $R_{ds}$ lowering (LRS, Fig. 4D right upper).

The Te SF-FET device enables a nonvolatile resistive memory with multilevel high and low resistance states. Figs. 4 E and F show the stable $V_p$ dependence of the resistance. The device has a high HRS/LRS resistance ratio nearly 100. After applying several pulsed voltages, the $I_{ds}$-$V_{ds}$ curves still remain linear but have different slopes (inset in Fig. 4F). Multilevel resistive states are achieved by pulsed gate voltages and are attributed to the moving and switching of multidomains. The continuous-variable resistive states could be used for multi-bit recording of information.

To summarize, we have discovered a room-temperature out-of-plane ferroelectricity in a nanowire of monoelemental material Te. A polar structure is detected through the STEM, resulting from a non-centrosymmetric lattice. PFM is used to probe the field-driven domain reversal and hysteresis behaviors. Based on the DFT simulation, it is concluded that the observed ferroelectric polarization can be attributed to the off-center ion displacements along the inclined edge surface of Te. Moreover, as a ferroelectric semiconductor, a Te nanowire possesses high mobility along the Te chain and out-of-plane ferroelectricity, which can be employed to fabricate an SF-FET. Such a device exhibits nonvolatile switching of resistance, multilevel resistive states and high on/off ratio. Our work expands the scope of ferroelectric materials and establishes a new platform for multi-bit zero-supply-voltage resistive memories.

**ACKNOWLEDGMENTS**

We thank Z. B. Hu and H. Ma for helpful discussions and C. L. Liu for experimental support. Funding: This work was supported in-part by the National Natural Science Foundation of China (Grant No. 11974304, 62004136, 11904250), and the Natural Science Foundation of Jiangsu Higher Education Institutions (20KJB140019), and also by Doctor of Entrepreneurship and Innovation in Jiangsu Province (No. (2020)30790), and also by Macau Science and Technology Development Fund (FDCT Grants 0106/2020/A3 and 0031/2021/ITP) and Macau University of Science and Technology (FRG-21-034-MISE). **Author contributions:** Y.C.J. conceived the idea and designed the experiments; Y.C.J. and Y.P.C. supervised the project and coordinated the collaboration; S. N. G. grew the Te crystals and fabricated the devices; S.N.G., Y.C.J., J.L.Z., and D.P.C. performed the measurements; Y.C.J. and H.B.Y. were responsible for the TEM measurements; J.Y.Z. and C.L.M. performed DFT calculations; Y.C.J., Y.P.C., R.Z., Y.P.Q., Z.P.W., L.W. and J.G. analyzed the data and interpreted the experimental results; J.L.Z., Y.C.J., J.Y.Z. and Y.P.Q. wrote the manuscript. All authors discussed the results and commented on the manuscript. **Competing interests:** The authors declare no competing interests. **Data and materials availability:** All data in the manuscript or the supplementary materials are available upon reasonable request.